\documentclass[aps,pra,twocolumn,preprintnumbers,groupedaddress,amsmath,amssymb,showkeys]{revtex4-1}
\usepackage{mathtools}
\usepackage{bm}
\usepackage{dsfont}
\usepackage{xcolor}

\newcommand{\pidx}[1]{{\mbox{\tiny $(#1)$}}}
\newcommand*{\placeholderdot}{\makebox[2ex]{$\textcolor{gray}{\bullet}$}}%

\makeatletter
\def\convertto#1#2{\strip@pt\dimexpr #2*65536/\number\dimexpr 1#1}
\makeatother

\usepackage[normalem]{ulem}

\begin{document}
	
	
	\title{Permanent Directional Heat Currents in Lattices of Optomechanical Resonators}
	
	
	\author{Zakari Denis}
	\affiliation{Universit\'e de Paris, Laboratoire Mat\'eriaux  et  Ph\'enom\`enes  Quantiques, CNRS, F-75013 Paris, France}
	\author{Alberto Biella}
	\affiliation{Universit\'e de Paris, Laboratoire Mat\'eriaux  et  Ph\'enom\`enes  Quantiques, CNRS, F-75013 Paris, France}
	\author{Ivan Favero}
	\affiliation{Universit\'e de Paris, Laboratoire Mat\'eriaux  et  Ph\'enom\`enes  Quantiques, CNRS, F-75013 Paris, France}
	\author{Cristiano Ciuti}
	\affiliation{Universit\'e de Paris, Laboratoire Mat\'eriaux  et  Ph\'enom\`enes  Quantiques, CNRS, F-75013 Paris, France}
	
	
	\date{\today}
	
	\begin{abstract}
		We study the phonon dynamics in lattices of optomechanical resonators where the mutually coupled photonic modes are coherently driven and the mechanical resonators are uncoupled and connected to independent thermal baths.
		We present a general procedure to obtain the effective Lindblad dynamics of the phononic modes for an arbitrary lattice geometry, where the light modes play the role of an effective reservoir that mediates the phonon nonequilibrium dynamics. We show how to stabilize stationary states exhibiting directional heat currents over arbitrary distance, despite the absence of thermal gradient and of direct coupling between the mechanical resonators.
	\end{abstract}
	
	
	\maketitle
	
	{\it Introduction ---} The emergence of persistent currents in many-body systems is tightly bound to fundamental concepts in classical and quantum physics. 
	In classical electrodynamics, any permanently magnetized object exhibits persistent electronic currents \cite{Griffiths1981}. 
	A conducting ring in the quantum coherent regime supports a permanent electric current when pierced by an external magnetic field \cite{Bouchiat1989}. 
	When pairing interactions are considered, a superconductor cooled below critical temperature displays persistent currents, and a constant magnetic field builds up through any continuous loop of the material \cite{Tinkham2004}. Systems with nontrivial topology can also give rise to persistent edge currents \cite{Rivas2017}.
	
	These manifestations of persistent currents involve two noticeable ingredients: (i) an external gauge field and (ii) the presence of a significant coherence extending over the entire sample \cite{Fazio2018}.
	Recently, it was shown that these ingredients are not strictly required, and that permanent currents in rings can instead be generated by {\it reservoir engineering} \cite{Fazio2018,Metelmann2015}, where specific many-body quantum states with properties of interest are stabilized \cite{Verstraete2009,Diehl2008}. More than a mere source of decoherence, the environment becomes then a tool to generate correlated phases, sometimes with no equilibrium counterpart \cite{Jin2016}.   
	In this context, the study of systems driven by nonlocal dissipators has emerged, notably in relation to nonreciprocal behaviors \cite{Metelmann2015}. In several nonreciprocal realizations, a direct coupling between two bosonic modes was engineered through a common ancillary degree of freedom \cite{Fang2017,Barzanjeh2018,Xu2019}. Very recently, the concept of engineered directionality was theoretically scaled up to extended lattices, by tailoring ancilla-assisted interactions \cite{Schmidt15,McDonald2018,Fazio2018,Verhagen2019}.
	
	Besides nonreciprocity, the coupling of independent mechanical modes to commonly shared optical modes was proposed to transport phonons between distant resonators \cite{Xuereb2014}, to model out-of-equilibrium quantum thermodynamics \cite{Xuereb2015}, and experimentally implemented to phase-lock adjacent \cite{Zhang2015} and distant \cite{Bagheri2013,GilSantos2017} mechanical resonators. Yet, many aspects of the nonlocal quantum dynamics of extended lattices in optomechanics remain to be explored despite their potentially uncommon features.
	
	In this work, we study analytically the effective dynamics of originally independent mechanical resonators coupled to extended lattices of driven-dissipative optical cavities. We express the general master equation of such a reservoir-coupled system and compare our predictions with a mean-field approach. Our study demonstrates that rings of lattices of optically coupled optomechanical resonators \cite{RevModPhysOM, Favero2009} can exhibit permanent whirling phonon currents. The latter are mediated by spatially correlated quantum fluctuations of the optical fields, in the absence of direct mechanical coupling, and triggered by proper tuning of the phase of the optical drive. The magnitude of the current is expressed analytically within a Born-Markov approximation, while the heat flow persists when mechanical resonators interact with independent thermal baths, over a wide range of temperatures.
	
	The existence of permanent phonon currents despite the absence of thermal gradient and of direct coupling between mechanical resonators is a novel phenomenon with no counterpart in models studied so far.
	
\begin{figure}[t!]
		\centering
		\includegraphics[width=\linewidth]{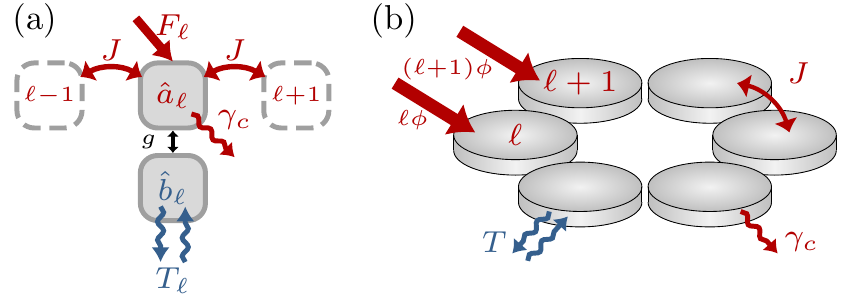}
		\caption{\label{fig:1}%
			(a) Schematic representation of a single optomechanical cell and its nearest-neighbor optical couplings. $\hat{a}_\ell$ ($\hat{b}_\ell$) is the optical (mechanical) mode of index $\ell$. (b) Ring of optomechanical disk resonators. Each site is optically driven with a phase that varies as $\ell\phi$, being $\ell$ the site number. Optical modes are coupled while mechanical ones are not.
		}
	\end{figure}

	{\it Generic model ---} 
	The  system under consideration consists of a network of $L$ optomechanical resonators whose optical modes are coherently driven by external laser fields. Neighboring cavities are optically coupled to one another, while mechanical modes are not. In a specific implementation with optomechanical disk resonators, optical modes are whispering gallery modes of adjacent resonators, while mechanical modes are radial breathing modes of individual disks \cite{Ding10}. Such resonators can be fabricated with ultralow site-to-site disorder \cite{GilSantos2017b}. One optomechanical cell is schematically illustrated in Fig.\,\ref{fig:1}\,(a).
	
	While in the following we focus on one-dimensional ($1$D) chains, here for the sake of generality, we consider an arbitrary network where the coupling between adjacent photonic modes is fully specified by a $L\times L$ adjacency matrix $\mathbf{A}$ where $A_{\ell\ell^\prime}=1$ if the sites $\ell$ and $\ell^\prime$ are coupled and $A_{\ell\ell^\prime}=0$ otherwise. In the frame rotating at the driving frequency $\omega_p$, the unitary part of the dynamics is described by the following Hamiltonian \cite{Ludwig13} ($\hbar = 1$):
	\begin{align}
	\hat{H}_\mathrm{tot} &= \sum_{\ell=1}^L \big[-\Delta_\ell \hat{a}_\ell^\dagger\hat{a}_\ell^{\mathstrut} + F_\ell^\star\hat{a}_\ell + F_\ell^{\mathstrut}\hat{a}_\ell^\dagger - g_\ell\hat{a}_\ell^\dagger\hat{a}_\ell^{\mathstrut}(\hat{b}_\ell^{\mathstrut} + \hat{b}_\ell^\dagger)\big]\nonumber\\
	&- \frac{J}{2}\sum_{\ell,\ell^\prime=1}^L A_{\ell\ell^\prime}\hat{a}_\ell^\dagger\hat{a}_{\ell^\prime}^{\mathstrut} + \sum_{\ell=1}^L \omega_m^{\pidx{\ell}}\hat{b}_\ell^\dagger\hat{b}_\ell^{\mathstrut},
	\label{eq:1}
	\end{align}
	where $\hat{a}_\ell$ and $\hat{b}_\ell$ are, respectively, the photonic and phononic annihilation operators of the $\ell$-th resonator, \smash{$\Delta_\ell = \omega_p - \omega_c^{\pidx{\ell}}$} denotes the detuning of the driving laser frequency with respect to the local bare cavity frequency $\omega_c^{(\ell)}$, $F_\ell$ is the (complex) amplitude of the coherent drive, $g_\ell$ is the optomechanical vacuum coupling rate and $J$ is the hopping rate between connected optical cavities. 
	Incoherent processes associated to local photon losses (at a rate $\gamma_c^{\pidx{\ell}}$) and phonon thermalization with their respective thermal baths (at a rate \smash{$\gamma_m^{\pidx{\ell}}$}) are taken into account by	means of a master equation for the system density matrix in the
	Lindblad form, which fully determines the system evolution,
	\begin{equation}
	\partial_t\hat{\rho}(t) = \mathcal{L}_\mathrm{tot}\hat{\rho}(t) \equiv -i[\hat{H}_\mathrm{tot},\hat{\rho}(t)] + \mathcal{D}_\mathrm{tot}\hat{\rho}(t),
	\label{eq:2}
	\end{equation}
	where 
	\begin{align}
	\mathcal{D}_\mathrm{tot}\hat{\rho} = \sum_{\ell=1}^L\Big\lbrace&\gamma_m^{\pidx{\ell}}\big[(\bar{n}_\ell + 1)\mathcal{D}[\hat{b}_\ell^{\mathstrut}]\hat{\rho} + \bar{n}_\ell\mathcal{D}[\hat{b}_\ell^\dagger]\hat{\rho}\big]\nonumber\\
	+&\gamma_c^{\pidx{\ell}}\mathcal{D}[\hat{a}_\ell]\hat{\rho}\Big\rbrace,
	\label{eq:3}
	\end{align}
	with $\mathcal{D}[\hat O]\hat{\rho}=\hat{O}\hat{\rho}\hat{O^\dagger} - \frac{1}{2} \lbrace\hat{O^\dagger}\hat{O},\hat{\rho} \rbrace$ and $\bar{n}_\ell$ the average number of thermal bosons due to the $\ell$-th thermal bath.

	Provided $\gamma_c \gg 2 g\langle\hat{a}^\dagger\hat{a}\rangle^{1/2}$, as we assume in what follows, optical fluctuations are negligibly affected by the mechanics and the coupled cavities can be regarded as an extended optical \emph{reservoir} which can safely be adiabatically eliminated in a wide range of the parameter space \footnote{When this is not the case the resulting effective master equation is dynamically unstable.}.
	By adjusting $J/\gamma_c$, one can tune the correlation length of the reservoir going from the uncoupled resonators local case ($J/\gamma_c \ll 1$) to that of a reservoir with resolved spectrum ($J/\gamma_c \gg 1$) \cite{Xuereb2014,Xuereb2015}.
	
	{\it Adiabatic elimination of an extended driven-dissipative reservoir ---}
	By splitting the fields into their mean-field value plus zero-mean fluctuations as $\hat{a}_\ell = \alpha_\ell + \hat{c}_\ell$ and $\hat{b}_\ell = \beta_\ell + \hat{d}_\ell$, we can expand the Hamiltonian and the dissipator to second order in the fluctuations around mean field,
	\begin{align}
	\hat{H}_\mathrm{tot}^\prime \simeq& \sum_{\ell=1}^L\Big[-\widetilde{\Delta}_\ell\hat{c}_\ell^\dagger\hat{c}_\ell^{\mathstrut}- \frac{J}{2}{\textstyle\sum_{\ell^\prime}}  A_{\ell\ell^\prime} \hat{c}_\ell^\dagger\hat{c}_{\ell^\prime}^{\mathstrut}+\hat{V}_\ell + \omega_m^{\pidx{\ell}}\hat{d}_\ell^\dagger\hat{d}_\ell^{\mathstrut}\Big],\label{eq:4}
	\end{align}
	where $\widetilde{\Delta}_\ell \approx \Delta_\ell + 2 g_\ell^2\lvert\alpha_\ell\rvert^2/\omega_m^{\pidx{\ell}}$ (for a high mechanical quality factor), $\hat{V}_\ell = (G_\ell^*\hat{c}_\ell^{\mathstrut}+G_\ell^{\mathstrut}\hat{c}_\ell^\dagger)(\hat{d}_\ell^{\mathstrut}+\hat{d}_\ell^\dagger)$, $G_\ell = g_\ell\alpha_\ell$,  and the mean fields $\lbrace\alpha_\ell, \beta_\ell\rbrace$ respect a self-consistency relation \footnote{Namely, $F_\ell - \tilde{\Delta}_\ell\alpha_\ell - i\frac{\gamma_c^{\pidx{\ell}}}{2}\alpha_\ell - \frac{J}{2}{\textstyle\sum_{\ell^\prime}}A_{\ell\ell^\prime}\alpha_{\ell^\prime} = 0$, and $\beta_\ell = \frac{g_\ell\lvert\alpha_\ell\rvert^2}{\omega_m^{\pidx{\ell}} - i\gamma_m^{\pidx{\ell}}/2}$. This insures that so long as the system remains dynamically stable ($\hat{c}_\ell(t\gg 1/\gamma_c) \approx 0$ and, thus, $\hat{a}_\ell(t\gg 1/\gamma_c) \approx \alpha_\ell$) the optical fluctuations have little memory on timescales larger than $1/\gamma_c$ and thus the single-body two-time correlation functions of the optical reservoir decay in time as those of a thermal bath at $T=0$.} that exactly cancels all linear terms in the fluctuation operators in the Hamiltonian. Both the amplitude and the phase of $G_\ell$ can be tuned through the driving. The dissipator remains that of Eq.~\eqref{eq:3} substituting $\hat{a}_\ell, \hat{b}_\ell \rightarrow \hat{c}_\ell, \hat{d}_\ell$. In this displaced frame, it becomes clear that finite-lived ($\tau_c=1/\gamma_c$) quantum optical fluctuations are not externally driven but can enter the reservoir from the mechanics through the now linear optomechanical coupling ($\hat{V}_\ell$) and can then be scattered back into some distant mechanical mode or be dissipated. We formalize this intuition hereafter by looking at the reduced dynamics of the mechanical degrees of freedom.
	
	Within the Born-Markov approximation, the lattice of optical cavities can be adiabatically eliminated (see Supplemental Material \footnote{See Supplemental Material, including Refs.~\cite{breuer-petruccione2007,Lebreuilly2016,Lebreuilly2017,RevModPhysQFL,Kronwald2013}, for further details.}) yielding the following effective Hamiltonian and dissipator for the mechanical modes:
	\begin{align}
	\hat{H}_m^{\mathrm{eff}} &= \sum_{\ell=1}^L\omega_m^{\pidx{\ell}}\hat{d}_\ell^\dagger\hat{d}_\ell^{\mathstrut} + \sum_{\ell,\ell^\prime=1}^L (\Omega_{\ell\ell^\prime}^\pidx{+}+\Omega_{\ell^\prime\ell}^\pidx{-})\hat{d}_\ell^\dagger\hat{d}_{\ell^\prime}^{\mathstrut},
	\label{eq:5}\\
	\mathcal{D}_m^{\mathrm{eff}}\hat{\rho}_m &= \sum_{\ell=1}^L\gamma_m^{\pidx{\ell}}\big[(\bar{n}_\ell + 1)\mathcal{D}[\hat{d}_\ell^{\mathstrut}]\hat{\rho}_m + \bar{n}_\ell \mathcal{D}[\hat{d}_\ell^{\dagger}]\hat{\rho}_m\big]\nonumber\\
	&+ \sum_{\ell=1}^L\big(\Gamma_{\ell}^\pidx{+} \mathcal{D}[\hat{\beta}_\ell^{\pidx{\downarrow}}]\hat{\rho}_m + \Gamma_{\ell}^\pidx{-} \mathcal{D}[\hat{\beta}_\ell^{\pidx{\uparrow}}]\hat{\rho}_m\big),
	\label{eq:6}
	\end{align}
	where $\pmb{\Omega}^\pidx{\pm} = \frac{1}{2i}\big(\mathbf{S}^\pidx{\pm} - \mathbf{S}^{\pidx{\pm}\dagger}\big)$ is the effective coherent coupling, determined by the spectrum of the extended reservoir:
	\begin{equation}
	S_{\ell\ell^\prime}^\pidx{\pm} = G_\ell^{\star} \bigg[\frac{i\mathbf{1}}{\pm\omega_m^\pidx{\ell^\prime}\mathbf{1} - \mathbf{B}}\bigg]_{\ell\ell^\prime} G_{\ell^\prime}^{\mathstrut},
	\label{eq:7}
	\end{equation}
	with $\mathbf{B} = -\frac{J}{2}\mathbf{A} - \mathrm{Diag}(\lbrace \widetilde{\Delta}_\ell+i\tfrac{\gamma_c^{\pidx{\ell}}}{2}\rbrace)$. The nonlocal dissipation rates are given by the eigenvalues of its Hermitian part, $\mathrm{Diag}(\lbrace\Gamma_{\ell}^\pidx{\pm}\rbrace) = \mathbf{U}^\pidx{\pm}(\mathbf{S}^\pidx{\pm} + \mathbf{S}^{\pidx{\pm}\dagger})\mathbf{U}^{\pidx{\pm}\dagger}$, where $\mathbf{U}^\pidx{\pm}$ are the associated diagonalizing unitary matrices. Finally, the nonlocal jump operators are defined as $\hat{\beta}_\ell^{\pidx{\downarrow}}=\sum_{\ell^\prime=1}^L U_{\ell\ell^\prime}^{\pidx{+}} \hat{d}_{\ell^\prime}^{\mathstrut}$, $\hat{\beta}_\ell^{\pidx{\uparrow}}=\sum_{\ell^\prime=1}^L U_{\ell\ell^\prime}^{\pidx{-}} \hat{d}_{\ell^\prime}^\dagger$ (note that in general  $\hat{\beta}_\ell^{\pidx{+}}\neq\hat{\beta}_\ell^{\pidx{-}\dagger}$). These results can be extended to continuous reservoirs and two-tone driven reservoirs generating multimode squeezing (see Supplemental Material \cite{Note3}).
	
	In the case of a finite 1D chain with nearest-neighbors coupling as henceforth considered, $\mathbf{B}$ takes the form of a tridiagonal matrix and has thus explicit inverse expressions \cite{Huang1997}. When $\lvert\Delta\pm\omega_m -i\gamma_c/2\rvert \lessapprox J/2$, off-diagonal elements have exponentially decreasing magnitudes, $\lvert S_{\ell,\ell+p}^\pidx{\pm}/S_{\ell,\ell}^\pidx{\pm}\rvert \sim (J/2)^p/\lvert\Delta\pm\omega_m -i\gamma_c/2\rvert^p$, so that the reservoir mainly couples neighboring mechanical modes, as expected from the finite lifetime of the optical fluctuations within the optical lattice. The range of the effective interaction can thus be selected by tuning $J/\gamma_c$. In contrast to previous works, where directional nearest-neighbors couplings at each edge $\langle \ell,\ell^\prime\rangle$ are engineered by having an independent ancillary mode coupled exclusively to the sites $\ell$ and $\ell^\prime$, we can simply rely on the finite lifetime of the mediating photon fluctuations to make the interaction short range when required.
	
	In this effective description, the lattice of cavities modifies the dynamics of the mechanical modes by adding coherent phonon-hopping processes between previously uncoupled mechanical modes and acting as a thermal bath for $L$ extended phononic modes $\{ \hat{\beta}_\ell^{\pidx{\downarrow}} \}$.

	\begin{figure}[t!]
		\centering
		\includegraphics[width=\linewidth]{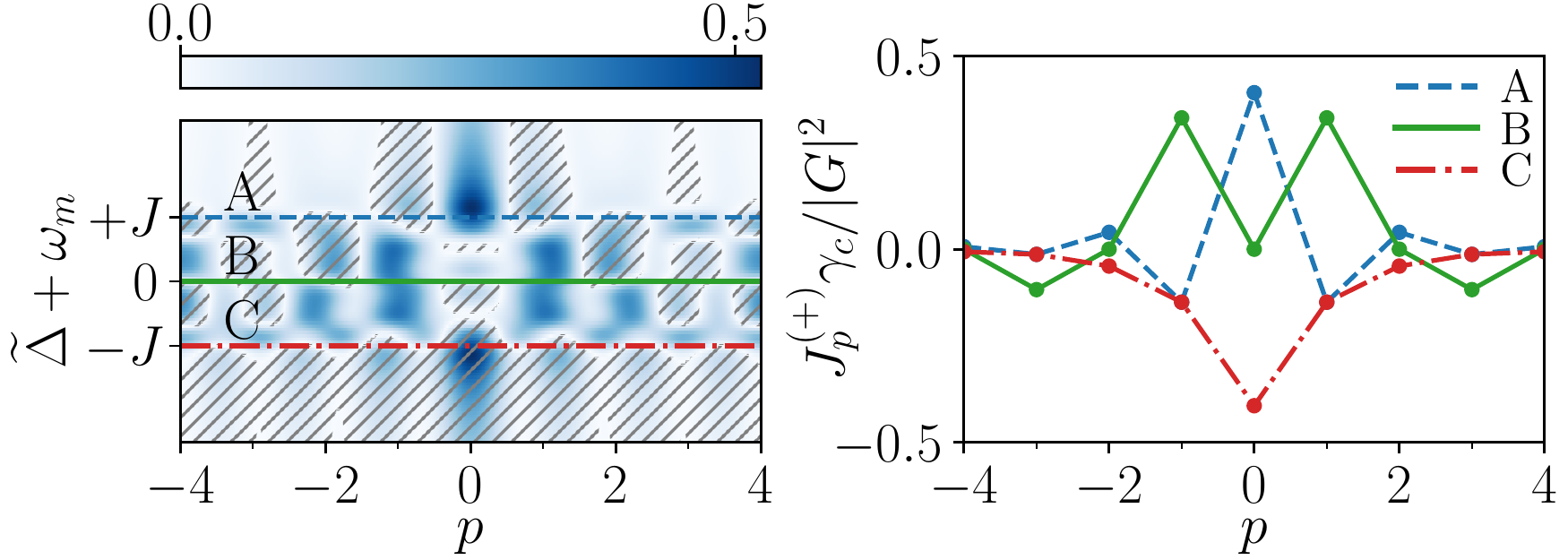}
		\caption{\label{fig:2}%
			Left panel: absolute value of the effective coherent coupling rate $J_p^{\pidx{+}}$ between mechanical sites as a function of their distance $p$ and the nonlinear detuning $\widetilde{\Delta}$. The hatched regions correspond to negative values. Three horizontal section cuts (labeled as A, B, C) are plotted in the right panel. Here $J/\gamma_c = 2$.
		}
	\end{figure}
	
\begin{figure}[t!]
		\centering
		\includegraphics[width=\linewidth]{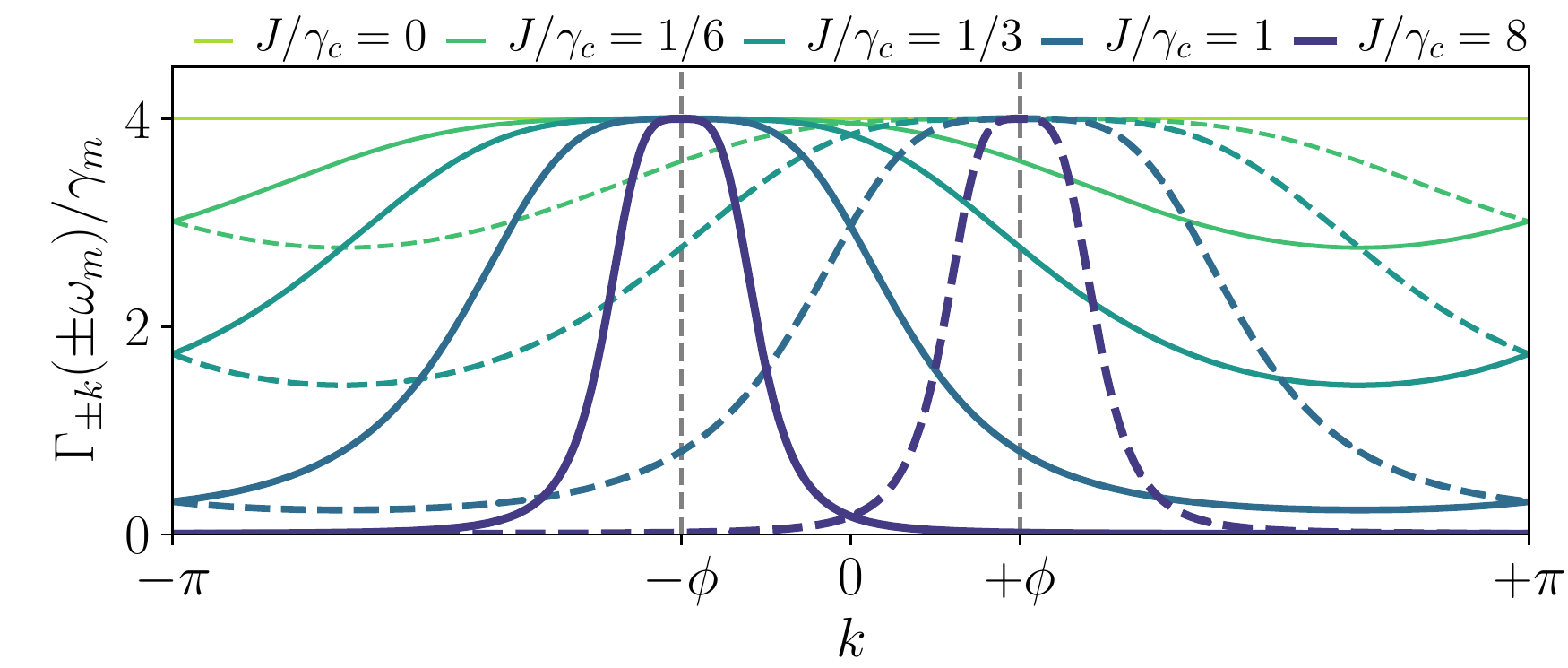}
		\caption{\label{fig:3}%
			Gain $\Gamma_{-k}(-\omega_m)$ (dashed) and loss $\Gamma_{k}(+\omega_m)$ rates induced by the engineered reservoir for various $J/\gamma_c$. $\widetilde{\Delta} = +\omega_m-J$ for the gain rate and $\widetilde{\Delta} = -\omega_m-J$ for the loss rate.
		}
	\end{figure}

	{\it Periodic 1D optomechanical lattice ---}
	\footnotetext[5]{A similar strategy has been adopted in Ref.~\cite{Hafezi2018} in optomechanical crystals where optical and mechanical modes are strongly hybridized and neighboring mechanical modes are directly coupled.}
	We now exploit the effective description derived above to study the emergence of persistent directional heat currents in an experimentally relevant model: a ring composed of $L$ sites.
	To this aim, the cavities are driven individually with the same intensity but with a site-dependent phase such that $|F_\ell|=F$ and $\mathrm{Arg}(F_\ell)=\ell\phi$ with $\phi=2\pi n/L$ and $n\in\mathbb{Z}$, which creates a homogeneous phase gradient around the ring \cite{Note5, Hafezi2018}. This situation is schematically illustrated in Fig.\,\ref{fig:1}\,(b). Following  Eqs. \eqref{eq:5} and \eqref{eq:6}, the unitary part of the mechanical effective dynamics is governed by:
	\begin{align}
	\hat{H}_m^{\mathrm{eff}} &= \sum_\ell(\omega_m+J_0^\pidx{+}+J_0^\pidx{-})\hat{d}_\ell^\dagger\hat{d}_\ell^{\mathstrut}\nonumber\\
	&+{\textstyle\sum_\pm}\sum_{\mathclap{1\leq p < L}} \frac{J_p^{\pidx{\pm}}}{2}\sum_\ell\big(\hat{d}_{\ell+p}^\dagger\hat{d}_\ell^{\mathstrut}e^{\mp i\phi\times p} + \mathrm{H.c.}\big),\label{eq:8}
	\end{align}
	where 
	\begin{equation}
	J_{p}^{\pidx{\pm}} = \sum_k\frac{e^{ikp}}{L}\frac{\lvert G\rvert^2(\pm\omega_m + \widetilde{\Delta} + J\cos k)}{(\pm\omega_m + \widetilde{\Delta} + J\cos k)^2+(\gamma_c/2)^2}
	\label{eq:9}
	\end{equation}
	is the real-valued amplitude of the effective complex coupling between $p$-distant modes. The second line involves two sets of directional couplings, noted by $\pm$. This can be understood from second order perturbation theory by examining the two mechanics-mechanics scattering processes having finite overlap $\langle f\rvert\hat{V}_{\ell+p}\hat{V}_\ell\lvert i\rangle$ and preserving the total energy: $\langle f\rvert G_{\ell+p}^\star\hat{c}_{\ell+p}\hat{d}_{\ell+p}^\dagger \times G_\ell\hat{c}_\ell^\dagger\hat{d}_\ell\lvert i\rangle$ and $\langle f\rvert G_{\ell+p}\hat{c}_{\ell+p}^\dagger\hat{d}_{\ell+p}^\dagger \times G_\ell^\star\hat{c}_\ell\hat{d}_\ell\lvert i\rangle$. The magnitude of each of these directional hopping channels, and thus the net effective flux of phonons, can be adjusted via the drive detuning $\Delta$. This dependence is complex in general, as shown in Fig.\,\ref{fig:2} for $J_p^{\pidx{+}}$. For this figure, as for all the following ones, parameters are $L=8$, $\phi = 2\pi/L$, $\lvert\alpha\rvert^2 = 100$, $g/\omega_m=2\cdot 10^{-3}$, $\gamma_c/\omega_m = 1\cdot 10^{-1}$, $\gamma_m/\omega_m = 1\cdot 10^{-3}$, and $\bar{n}=100$.
	The incoherent part of the effective dynamics is given by Eq.\,\eqref{eq:6} by substituting 
	$\gamma_m^{(\ell)},\bar{n}_\ell \to \gamma_m, \bar{n}$;
	$\hat{\beta}_\ell^{\pidx{\downarrow}}, \hat{\beta}_\ell^{\pidx{\uparrow}} \to \tilde{d}_k^{\mathstrut}, \tilde{d}_{-k}^\dagger$;
	$\Gamma_\ell^{(+)},\Gamma_\ell^{(-)} \to \Gamma_k(+\omega_m),\Gamma_k(-\omega_m)$.
	The Fourier modes being defined as $\tilde{d}_k = \tfrac{1}{\sqrt{L}}{\textstyle\sum_\ell} e^{-ik\ell}\hat{d}_\ell$ with $k \in \lbrace n\times2\pi/L\rbrace_{n=0}^{L-1}$ and
	\begin{equation}
	\Gamma_{k}(\omega) = \frac{\lvert G\rvert^2\gamma_c}{(\omega + J\cos(k+\phi)+\widetilde{\Delta})^2+(\gamma_c/2)^2}.
	\label{eq:10}
	\end{equation}
	In contrast to the single resonator case \cite{Marquardt2007}, our system has $L$ Stokes sidebands at $\widetilde{\Delta}^\pidx{-}_k = \omega_m - J\cos(k-\phi)$ and $L$ anti-Stokes sidebands at $\widetilde{\Delta}^\pidx{+}_k = -\omega_m - J\cos(k-\phi)$, that can be employed to respectively amplify or cool collective mechanical modes. In Fig.~\ref{fig:3} we show the $k$-space asymmetry between the incoherent gain and loss rates for $\phi\neq0$ around the lowest Stokes and anti-Stokes sidebands. Depending on the detuning, the engineered optical reservoir acts onto the system either by absorbing collective excitations with pseudomomentum $k\sim-\phi$ (jump operator $\tilde{d}_k$) or by creating excitations with opposite momentum $k\sim+\phi$ (jump operator $\tilde{d}_{-k}^\dagger$). Let us stress that this is not the result of the optical driving being at resonance with any particular $k$ mode as it holds when the dissipation rate is of the order of the width of the optical lattice's spectrum ($J \sim \gamma_c$). In such a regime, the concept of resonance has no longer any operative meaning.

	Let us now investigate the steady state properties of this effective model by diagonalizing the Liouvillian in the Fourier mode basis as $\hat{H}_m^{\mathrm{eff}} = {\textstyle\sum_k} \omega_k \tilde{d}_k^\dagger\tilde{d}_k^{\mathstrut}$ and $\mathcal{D}_m^{\mathrm{eff}}\hat{\rho}_m = {\textstyle\sum_k}\big(\Gamma_k^{\pidx{\downarrow}}\mathcal{D}[\tilde{d}_k^{\mathstrut}]\hat{\rho}_m+ \Gamma_k^{\pidx{\uparrow}}\mathcal{D}[\tilde{d}_{-k}^{\dagger}]\hat{\rho}_m\big)$, with
	\begin{align}
	\omega_k &= \omega_m + {\textstyle\sum_\pm}\frac{\lvert G\rvert^2(\pm\omega_m + \widetilde{\Delta} + J\cos(k\pm\phi))}{(\pm\omega_m + \widetilde{\Delta} + J\cos(k\pm\phi))^2 + (\gamma_c/2)^2},\label{eq:11}\\
	\Gamma_k^{\pidx{\downarrow}} &= \gamma_m(\bar{n}+1) + \Gamma_k(+\omega_m)
	\;;\;\Gamma_k^{\pidx{\uparrow}} = \gamma_m \bar{n} + \Gamma_k(-\omega_m).\label{eq:12}
	\end{align}

	As can be seen in Eqs.\,\eqref{eq:11} and \eqref{eq:12}, both the unitary and the dissipative parts of the Liouvillian are no longer even in $k$ space for finite $\phi$, as a result of having explicitly broken the parity symmetry of the coupling to the reservoir. In particular, with a driving laser operated around the lowest anti-Stokes sideband ($\widetilde{\Delta} \approx -J-\omega_m$) and within the resolved sideband regime ($\gamma_c \ll \omega_m$), to first order in $J/\gamma_c$ the system has a noneven dispersion relation of the form $\omega_k = \mathrm{cst.} + 4(\lvert G\rvert/\gamma_c)^2 J\cos(k+\phi)$ with a ground state at finite momentum $k_\mathrm{GS}=+\phi$.
	
\begin{figure}[t!]\centering
		\includegraphics[width=1.0\linewidth]{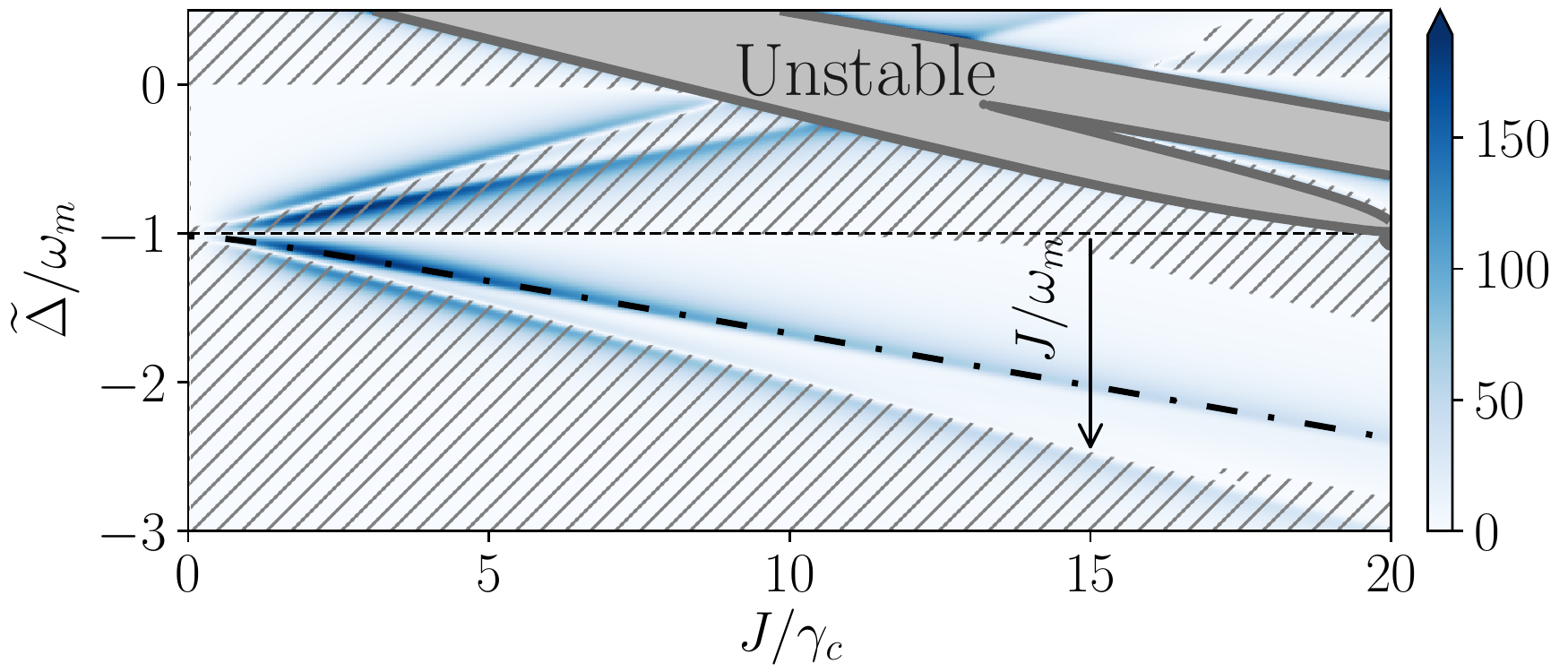}
		\caption{\label{fig:4}%
			Absolute value of the predicted net permanent heat current $\lvert Q_C\rvert$ in units of $\omega_m\times\gamma_m$ around a ring of cavity-coupled optomechanical resonators as a function of the intercavity coupling and the detuning. The hatched regions correspond to negative values. The system is unstable in the gray region.
		}
	\end{figure}

	{\it Permanent cavity-mediated directional heat current ---}
	The optical mean-field phase gradient yields a permanent directional heat flow around the ring of disks. Indeed, as discussed in the Supplemental Material \cite{Note3}, the continuity equation satisfied by the phonon number operator $i[\hat{H}_m^\mathrm{eff}, \hat{d}_\ell^\dagger\hat{d}_\ell^{\mathstrut}] = - \sum_p(\hat{\jmath}_{\ell\rightarrow\ell-p} + \hat{\jmath}_{\ell\rightarrow\ell+p})$, with $\hat{\jmath}_{\ell\rightarrow\ell+p} = -{\textstyle\sum_{\pm}} \frac{J_p^{\pidx{\pm}}}{2i}(\hat{d}_{\ell+p}^\dagger\hat{d}_{\ell}^{\mathstrut} e^{\mp i\phi p} - \mathrm{H.c.})$, induces the following definition for a net circulating current operator: $\hat{\jmath}_{C} = \sum_{\ell=1}^L\sum_{1\leq p<\lceil L/2\rceil}\hat{\jmath}_{\ell\rightarrow\ell+p}$.
	In $k$ space, it reads: 
	\begin{equation}
	\hat{\jmath}_{C}  
	= -\sum_k\sum_{1\leq p<\lceil L/2\rceil} \sum_{\pm} J_p^{\pidx{\pm}}\sin(p(k\pm\phi))\tilde{d}_k^\dagger\tilde{d}_k^{\mathstrut}.
	\label{eq:13} 
	\end{equation}
	The expectation value of this operator can be determined experimentally by measuring the thermal populations $\langle\tilde{d}_k^\dagger\tilde{d}_k^{\mathstrut}\rangle$, for example, via the mechanical noise spectrum around the $L$ collective mechanical frequencies $\omega_k$ measured at the output of some local resonator. For example in optomechanical disk resonators a secondary optical mode, such as a higher-order whispering gallery mode of the disk, could be used for that purpose.
	For our effective model, we get:
	\begin{equation}
	\langle\hat{\jmath}_C\rangle_\mathrm{ss} = -\sum_k\frac{\sum_{1\leq p<\lceil L/2\rceil} {\textstyle\sum_{\pm}} J_p^{\pidx{\pm}}\sin(p(k\pm\phi))}{\Gamma_k^{\pidx{\downarrow}}/\Gamma_{-k}^{\pidx{\uparrow}}-1}.
	\label{eq:14}
	\end{equation}
	The net permanent heat current whirling around the ring is thus simply $Q_C = \omega_m\langle\hat{\jmath}_C\rangle_\mathrm{ss}$. The amount of this heat transported over a phonon lifetime is shown in single phonon energy units in Fig.\,\ref{fig:4} as a function of $\widetilde{\Delta}/\omega_m$ and $J/\gamma_c$. Its sign (propagation direction) depends crucially on the detuning. Indeed, the effective coherent coupling can be regarded as an optical spring effect in $k$ space and, as such, it changes sign when crossing a sideband.

\begin{figure}[t!]\centering
		\includegraphics[width=\linewidth]{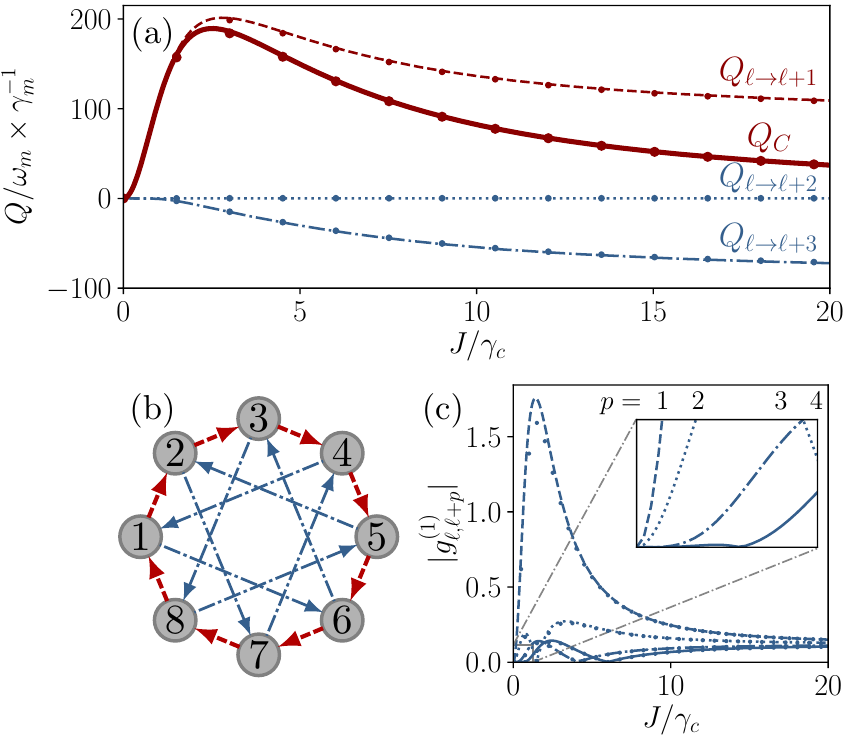}
		\caption{\label{fig:5}%
			(a) Contributions $Q_{\ell\rightarrow\ell+p} = \omega_m\sum_\ell\langle\hat{\jmath}_{\ell\rightarrow\ell+p}\rangle$ and net directional heat flow $Q_C = \sum_{p<\lceil L/2\rceil}Q_{\ell\rightarrow\ell+p}$ along the dash-dotted line of Fig.\,\ref{fig:4} as predicted by our effective theory (lines) and mean field (circles). (b) Sketch of the two leading contributions in (a). (c) Gradual triggering of off-diagonal coherence $g^{(1)}_{\ell,\ell+p} = \langle\hat{d}_\ell^\dagger\hat{d}_{\ell+p}\rangle / (\langle\hat{d}_\ell^\dagger\hat{d}_{\ell}\rangle\langle\hat{d}_{\ell+p}^\dagger\hat{d}_{\ell+p}\rangle)^{1/2}$ along the lowest anti-Stokes sideband $\widetilde{\Delta} = -J - \omega_m$.
		}
	\end{figure}

	In Fig.\,\ref{fig:5}\,(a), we show the behavior of contributions $Q_{\ell\rightarrow\ell+p} = \omega_m\sum_\ell\langle\hat{\jmath}_{\ell\rightarrow\ell+p}\rangle$ to the total flow 
	as a function of $J$ when the detuning $\widetilde{\Delta}$ is adjusted to follow its maximum (dash-dotted line of Fig.\,\ref{fig:4}). Interestingly, $Q_C$ is nonmonotonic in $J/\gamma_c$.	 For $J \lesssim \gamma_c$, optical fluctuation quanta mediating the heat transport are short lived ($\tau_c \lesssim 1/J$) and are thus destroyed before reaching sites farther than their nearest neighbors. This implies that the only sizable contribution is that flowing by local steps in the clockwise direction. Conversely, for $J \gtrsim \gamma_c$, optical fluctuation quanta can be scattered farther across the optical lattice before being destroyed by the cavity losses and the permanent heat flow is supported on supplementary directed graphs (see Fig.\,\ref{fig:5}\,(b)). In this case, a nonlocal anticlockwise flow contributes to the nonmonotonic dependence on $J/\gamma_c$ of the net current. Fig.\,\ref{fig:5}\,(c) shows how longer-range correlations get gradually triggered as the $J/\gamma_c$ ratio is increased following the lowest anti-Stokes sideband (see arrow in Figure~\ref{fig:4}).

	{\it Conclusion ---}
	We have studied the emergence of spatial correlations and permanent directional heat currents across lattices of optomechanical resonators whose mechanical modes are originally uncoupled. In our picture, quantum fluctuations of the optical fields mediate effective long-range interactions between mechanical sites of both coherent and dissipative nature, whose range is tunable via the correlation length of the reservoir.
	A remarkable feature is the possibility to flow arbitrary phonon streams in directions and topologies that seem to contradict common thermodynamic intuition, for example, a permanent phonon heat flow can be generated in the absence of thermal gradient.
	
	More generally, our investigation provides a first instance of a broader class of physical situations for which a weak coupling to an extended reservoir suffices to alter dramatically the fate of an initially trivial set of independent modes. The here presented effective description introduces an analytical tool for understanding quantum systems interacting via extended close-to-Markovian reservoirs, a realm yet to be fully explored.
	
	We thank D. Rossini for discussions. This work was supported by ERC via Consolidator Grants NOMLI No. 770933 and CORPHO No. 616233, and by ANR via the project UNIQ.

	\bibliography{references}
	
	\widetext
	\clearpage
	\begin{center}
		\textbf{\large Supplemental Material: Permanent directional heat currents in lattices of optomechanical resonators}
	\end{center}
	\setcounter{equation}{0}
	\setcounter{figure}{0}
	\setcounter{table}{0}
	\setcounter{page}{1}
	\makeatletter
	\renewcommand{\theequation}{S\arabic{equation}}
	\renewcommand{\thefigure}{S\arabic{figure}}
	
	\section{Adiabatic elimination of extended driven-dissipative reservoirs\label{app:A}}
	\subsection{Discrete reservoir}
	Let us consider some generic \emph{system} $S$ whose dynamics is governed by a Liouvillian $\mathcal{L}_S$ and a \emph{reservoir} $R$ described by a Liouvillian $\mathcal{L}_R$. The corresponding two sets of degrees of freedom are weakly coupled via a Hamiltonian term of the form $\hat{V} = \lambda \sum_i\hat{R}_i\otimes\hat{S}_i$ where $\lambda$ is a scale bookmark and where $\lbrace\hat{R}_i\rbrace_i$ and $\lbrace\hat{S}_i\rbrace_i$ act respectively on the reservoir and on the system Hilbert spaces. We associate sets of ladder operators $\lbrace\hat{a}_i\rbrace_i$ and $\lbrace\hat{b}_i\rbrace_i$ to the reservoir and the system, respectively. The vacuum is then displaced to a stable mean-field solution $\lbrace\alpha_i,\beta_i\rbrace_i$ towards which the reservoir is driven ($\hat{\rho} \mapsto \hat{D}^\dagger\hat{\rho}\hat{D}$, $\mathcal{L}\placeholderdot \mapsto \hat{D}\mathcal{L} \hat{D}^\dagger\placeholderdot$ with $\hat{D} = \exp(\alpha^i\hat{a}_i+\beta^i\hat{b}_i - \mathrm{H.c.})$) and the resulting Liouvillians are expanded to second order in the reservoir's ladder operators. By construction, the reservoir becomes thermal-like, in the sense that all its correlation functions decay exponentially in time, and can be traced-out by means of the Born-Markov procedure \cite{breuer-petruccione2007}. We first perform a Born approximation, by assuming that the state of the reservoir and the system, initially $\hat{\rho}(t_0) = \hat{\rho}_\mathrm{R}(t_0)\otimes\hat{\rho}_\mathrm{S}(t_0)$, remains separable upon time-evolution $\hat{\rho}(t) \approx \hat{\rho}_\mathrm{R}(t)\otimes\hat{\rho}_\mathrm{S}(t)$. This yields first the following non-Markovian master equation for the reduced density matrix of the system $\hat{\rho}_\mathrm{S}(t) = \mathrm{Tr}_\mathrm{R}[\hat{\rho}(t)]$~\footnote{One gets exactly the same expression by using the Nakajima-Zwanzig projective method within the Born approximation (cf. \S 9.1.2 of \cite{breuer-petruccione2007}) as in \cite{Xuereb2015,Lebreuilly2016,Lebreuilly2017}}
	\begin{align}
	\partial_t\hat{\rho}_\mathrm{S}(t) &= \mathrm{Tr}_\mathrm{R}[\mathcal{L}_\mathrm{R}\hat{\rho}(t) + \mathcal{L}_\mathrm{S}\hat{\rho}(t)-i[\hat{V},\hat{\rho}(t)]] \equiv \mathcal{L}_\mathrm{S}\hat{\rho}_\mathrm{S}(t) + \delta\mathcal{L}_\mathrm{S}\hat{\rho}_\mathrm{S}(t),\\ \delta\mathcal{L}_\mathrm{S}\hat{\rho}(t) &= -\lambda^2\sum_{ij} \int_{0}^{t-t_0}\mathrm{d}\tau \big\lbrace\langle\hat{R}_i(t-t_0)\hat{R}_j(t-t_0-\tau)\rangle_R[\hat{S}_{i},e^{\mathcal{L}_\mathrm{S}\tau}(\hat{S}_{j}\hat{\rho}_\mathrm{S}(t-\tau))]+\mathrm{H.c.}\big\rbrace\nonumber\\
	&\xrightarrow{t_0\rightarrow-\infty} -\lambda^2\sum_{ij} \int_{\mathbb{R}_+}\mathrm{d}\tau \big\lbrace\mathcal{G}_{ij}(\tau)[\hat{S}_{i},e^{\mathcal{L}_\mathrm{S}\tau}(\hat{S}_{j}\hat{\rho}_\mathrm{S}(t-\tau))]+\mathrm{H.c.}\big\rbrace,
	\label{eq:a1}
	\end{align}
	where it was assumed for simplicity that $\langle\hat{V}\rangle(t\rightarrow+\infty) = 0$ (e.g. $\hat{V}$ normal-ordered in $\lbrace\hat{a}_i\rbrace_i$) and where
	\begin{equation}
	\mathcal{G}_{ij}(\tau) = \lim_{t\rightarrow+\infty}\langle\hat{R}_i(t)\hat{R}_j(t-\tau)\rangle_R.
	\end{equation}
	The limit in the last line is taken as we only aim at describing the dynamics at a timescale larger than the relaxation time of the reservoir.
	For consistency, we also assume that the system was weakly coupled to its environment so that $\lambda^2\exp(\mathcal{L}_S\tau_c)\placeholderdot \approx \lambda^2\exp(-i[\hat{H}_S,\placeholderdot]\tau_c)$. We then perform a Markov approximation by neglecting the effect of the reservoir on the system on times of the order of the reservoir's correlation time $\tau_c$, yielding
	\begin{align}
	\delta\mathcal{L}_\mathrm{S}\hat{\rho}(t) \approx -\lambda^2\sum_{ij} \int_{\mathbb{R}_+}\mathrm{d}\tau \big\lbrace\mathcal{G}_{ij}(\tau)[\hat{S}_{i},e^{-i\hat{H}_\mathrm{S}\tau}\hat{S}_{j}e^{+i\hat{H}_\mathrm{S}\tau}\hat{\rho}_\mathrm{S}(t)]+\mathrm{H.c.}\big\rbrace.
	\label{eq:a3}
	\end{align}
	By decomposing the coupling operators in the eigenoperator basis of $\hat{H}_S$ as $\hat{S}_i = \sum_{\alpha} \hat{s}_i(\omega_\alpha)$, where the $\lbrace\hat{s}_i(\omega)\rbrace_i$ are such that $[\hat{H}_S,\hat{s}_i(\omega)] = -\omega \hat{s}_i(\omega)$ and $\lbrace\omega_\alpha\rbrace_\alpha$ denotes the set of all possible transition energies between eigenstates of the system Hamiltonian $\hat{H}_S$, and, by neglecting counter-propagating terms (consistent with the assumption made in the previous paragraph), one finally gets
	\begin{equation}
	\delta\mathcal{L}_\mathrm{S}\hat{\rho}(t) \approx -\lambda^2 \sum_{i,j,\alpha}S_{ij}^{\pidx{\alpha}} \big[\hat{s}_i^\dagger(\omega_\alpha), \big[\hat{s}_j^{\mathstrut}(\omega_\alpha)\hat{\rho}_S(t)\big]\big] + \text{H.c.}
	\label{eq:a4}
	\end{equation}
	with $S_{ij}^{(\alpha)} = \int_{\mathbb{R}_+}\mathrm{d}\tau \mathcal{G}_{ij}(\tau) e^{i\omega_\alpha\tau}$. Its Lindblad form can be made explicit by identifying the Hermitian and anti-Hermitian components of the reservoir spectrum
	\begin{gather}
	\mathbf{S}^\pidx{\alpha} + \mathbf{S}^{\pidx{\alpha}\dagger} = \mathbf{U}^{\pidx{\alpha}\dagger} \mathrm{Diag}(\lbrace\Gamma_{i}^{\pidx{\alpha}}\rbrace) \mathbf{U}^{\pidx{\alpha}},\quad \text{with}\quad \mathbf{U}^{\pidx{\alpha}\dagger} = \mathbf{U}^{\pidx{\alpha}-1}\\
	\pmb{\Omega}^{\pidx{\alpha}} = \frac{1}{2i}\big(\mathbf{S}^{\pidx{\alpha}} - \mathbf{S}^{\pidx{\alpha}\dagger}\big),
	\end{gather}
	in terms of which Eq.\,\eqref{eq:a4} reads:
	\begin{align}
	\delta\mathcal{L}_\mathrm{S} &= \lambda^2\big\lbrace-i[{\textstyle\sum_{ij\alpha}} \Omega_{ij}^\pidx{\alpha}\hat{s}_i^\dagger\hat{s}_j^{\mathstrut}, \placeholderdot] + {\textstyle\sum_i}\Gamma_{i}^\pidx{\alpha} \mathcal{D}[U_{\;\; i\;\;}^{\pidx{\alpha}j}\hat{s}_j]\big\rbrace.
	\end{align}
	Assuming that the reservoir fluctuations remain Gaussian, for any choice of $\lbrace\hat{R}_i\rbrace_i$, $\mathbf{S}^{\pidx{\alpha}}$ can be computed from the covariance matrix:
	\begin{equation}
	\mathbf{C}(\tau\geq 0) = \big\langle\hat{\bm{A}}(\tau)\;\hat{\bm{A}}^T(0)\big\rangle = e^{-i\mathbf{B}\tau}\mathbf{C}(0),
	\end{equation}
	where $\hat{\bm{A}} = [\ldots,\hat{a}_i^{\mathstrut}, \hat{a}_i^\dagger,\ldots]^T$ and $\mathbf{B}$ is the Bogoliubov operator defined by the Bogoliubov-like equation $i\partial_t \hat{\bm{A}} = \mathbf{B}\hat{\bm{A}}$. In particular, for some generic linear form $\hat{R}_i = t_i^\star\hat{a}_i + \mathrm{H.c.}$ as in the main text, one obtains:
	\begin{equation}
	\mathbf{S}^{\pidx{\alpha}} = \mathbf{T}^T\frac{i\mathbf{1}}{\omega_\alpha\mathbf{1} - \mathbf{B}}\mathbf{C}(0)\mathbf{T},
	\end{equation}
	where $\mathbf{T}$ is given by the direct matrix sum $\mathbf{T} = \mathrm{Diag}\big(\ldots,\begin{bsmallmatrix}t_i^\star \\t_i\end{bsmallmatrix},\ldots\big)$. In the main text, no squeezing of the reservoir's fluctuations on top of mean-field was considered so that $\bm{\mathcal{G}}(\tau) = \mathbf{T}^T \mathbf{C}(\tau) \mathbf{T} = \mathbf{T}^T (\mathbf{C}'(\tau)\otimes\begin{bsmallmatrix}
	0 & 1\\
	0 & 0
	\end{bsmallmatrix}) \mathbf{T} = (\bm{t}\,\bm{t}^\dagger) \odot \mathbf{C}'(\tau) = (\bm{t}\,\bm{t}^\dagger) \odot e^{-i\mathbf{B}'\tau}\mathbf{C}'(0)$, with $C_{ij}^\prime(0) = \delta_{ij}$, and then $S_{ij}^{\pidx{\alpha}} = t_i^\star t_j\big[\frac{i}{\omega_\alpha\mathbf{1} - \mathbf{B}'}\big]_{i,j}$, where the simpler Bogoliubov operator was such that $i\partial_t \hat{a}_i = B_{ij}^\prime\hat{a}_j$.
	
	\subsection{Continuous reservoir}
	
	The same procedure can be applied to continuous reservoirs. For example, let us consider the case of some translational-invariant reservoir defined by the squeezed Gaussian fluctuations of a free condensate around its mean-field solutions:
	\begin{equation}
	\hat{H} = \int \mathrm{d}\mathbf{r}\mathrm{d}\mathbf{r}^\prime\hat{\bm{\Psi}}^\dagger(\mathbf{r}) \mathbf{H}(\mathbf{r}-\mathbf{r}^\prime)\hat{\bm{\Psi}}(\mathbf{r}^\prime)
	\;,\quad\mathcal{D} = \int\mathrm{d}\mathbf{r} \gamma_c\mathcal{D}[\hat{\psi}(\mathbf{r})],
	\end{equation}
	where the modes $\hat{\bm{\Psi}}(\mathbf{r}) = [\hat{\psi}(\mathbf{r}),\; \hat{\psi}^\dagger(\mathbf{r})]^T$ satisfy some Bogoliubov equation $i\partial_t\hat{\bm{\Psi}}(\mathbf{r}) = \int\mathrm{d}\mathbf{r}^\prime\mathbf{B}(\mathbf{r}-\mathbf{r}^\prime)\hat{\bm{\Psi}}(\mathbf{r}^\prime)$ \cite{RevModPhysQFL}, where the Bogoliubov operator typically carries some dependence on the mean fields accounting for the nonlinearity of the model. This continuous set of degrees of freedom is put in contact with some discrete set of mechanical modes via an interaction Hamiltonian
	\begin{equation}
	\hat{V} = \lambda \sum_i\hat{R}(\mathbf{r}_i) \otimes \hat{S}_i,
	\end{equation}
	with some general choice of local coupling $\hat{R}(\mathbf{r}) = \bm{t}^T(\mathbf{r}) \cdot \hat{\bm{\Psi}}(\mathbf{r})$.
	
	Under the above-discussed approximations, the system's effective master equation has the same expression as in the previous subsection, the only difference being the expression of the reservoir spectrum $S_{ij}^{\pidx{\alpha}} = \int_{\mathbb{R}_+}\mathrm{d}\tau e^{i\omega_\alpha\tau} \mathcal{G}(\mathbf{r}_i,\mathbf{r}_j;\tau)$, which here takes the form of the following convolution:
	\begin{align}
	S_{ij}^{\pidx{\alpha}} &= \bm{t}^T(\mathbf{r}_i)\widetilde{\mathbf{C}}(\mathbf{r}_i-\mathbf{r}_j;\omega_\alpha)\bm{t}(\mathbf{r}_j),\\
	\widetilde{\mathbf{C}}(\mathbf{r}_i-\mathbf{r}_j;\omega) &= \int\mathrm{d}\mathbf{r}\int\frac{\mathrm{d}\mathbf{k}}{(2\pi)^d} \frac{ie^{i\mathbf{k}\cdot(\mathbf{r}_i-\mathbf{r}_j-\mathbf{r})}}{\omega\mathbf{1} - \widetilde{\mathbf{B}}(\mathbf{k})}\mathbf{C}(\mathbf{r};0).
	\end{align}
	The covariance $\mathbf{C}(\mathbf{r};0) = \langle\hat{\bm{\Psi}}(\mathbf{r}) \hat{\bm{\Psi}}^T(\mathbf{0})\rangle$ is to be evaluated from the steady-state mean-field solution and $\widetilde{\mathbf{B}}(\mathbf{k}) = \int\mathrm{d}\mathbf{r}e^{-i\mathbf{k}\cdot\mathbf{r}} \mathbf{B}(\mathbf{r})$.
	
	\section{Benchmarking the effective description}
	
	In order to benchmark our effective description, we compute the exact steady-state covariance matrix of both optical and mechanical fluctuations for the linearised model described by Eq.~\eqref{eq:4} and extract the exact mean-field single-particle density matrix $\sigma_{mn} = \langle\hat{d}_m^\dagger\hat{d}^{\mathstrut}_n\rangle_\mathrm{ss}$ as given by:
	\begin{equation}
	\sigma^{\mathrm{MF}}_{\ell,\ell^\prime} = \lim_{t\rightarrow+\infty}\big\langle\hat{\bm{\phi}}(t)\;\hat{\bm{\phi}}^T(t)\big\rangle_{L+2\ell,L+2\ell^\prime-1}\;,
	\label{eq:V13}
	\end{equation}
	where $\hat{\bm{\phi}}(t) = [\hat{c}_1^{\mathstrut}(t), \hat{c}_1^\dagger(t), \ldots, \hat{d}_1^{\mathstrut}(t), \hat{d}_1^\dagger(t),\ldots]^T$, and compare it to the single-particle density matrix of the effective description given explicitly by
	\begin{equation}
	\sigma_{\ell\ell^\prime}^\mathrm{eff} = \frac{1}{L}\sum_{k}\frac{e^{-ik(\ell-\ell^\prime)}}{\Gamma_k^{\pidx{\downarrow}}/\Gamma_{-k}^{\pidx{\uparrow}}-1}
	\label{eq:V14}
	\end{equation}
	by computing the error $\delta = \lVert\sigma^\mathrm{eff}-\sigma^\mathrm{MF}\rVert_2 / \lVert(\sigma^\mathrm{eff}+\sigma^\mathrm{MF})/2\rVert_2$.
	
\begin{figure}[ht!]\centering
		\includegraphics[width=.9\linewidth]{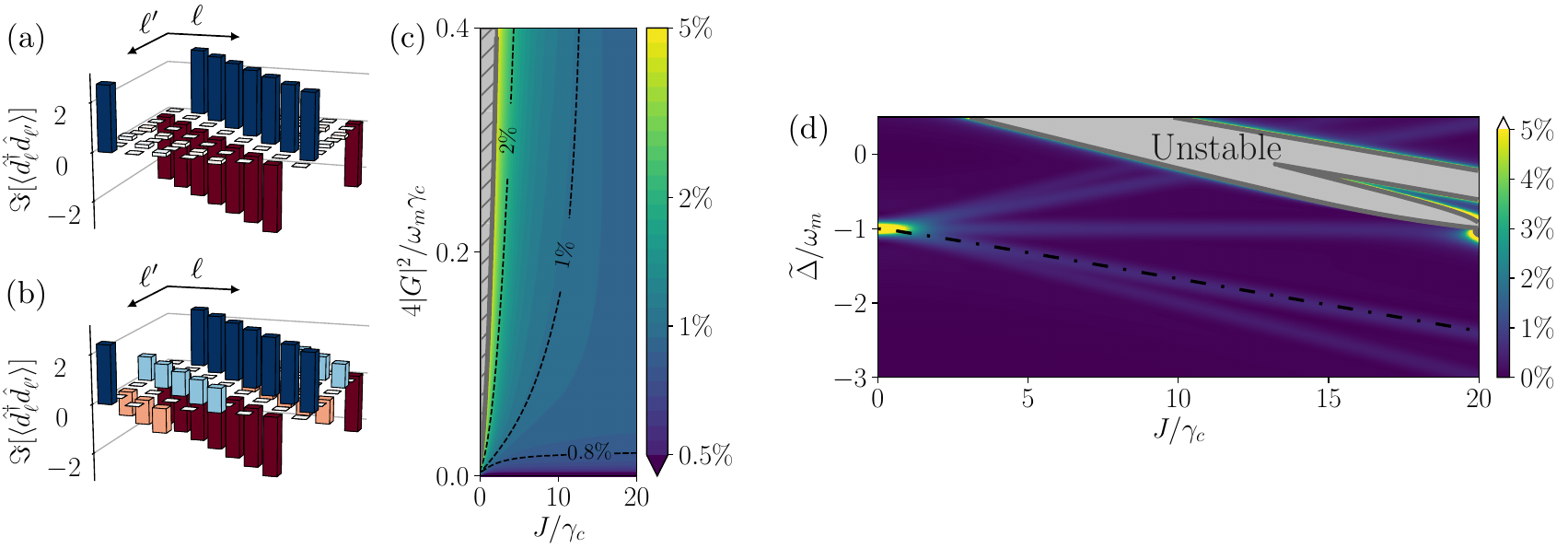}
		\caption{\label{fig:S1}%
			Imaginary part of the steady-state single-particle density matrix predicted by the effective theory as given by Eq.\,\ref{eq:V14} for $L=8$, $\phi = 2 \times 2\pi/L$, $\lvert\alpha\rvert^2 = 100$, $\widetilde{\Delta} = -J-\omega_m$, $g/\omega_m=1\cdot 10^{-2}$, $\gamma_c/\omega_m = 1\cdot 10^{-1}$, $\gamma_m/\omega_m=1\cdot 10^{-3}$, $\bar{n}=10$ and $J/\gamma_c = 1$ (a) and $J/\gamma_c = 5$ (b). Relative errors $\delta$ are $1.0\%$ (a) and $0.7\%$ (b). (c) $\delta$ as a function of the effective optomechanical coupling and the inter-cavity coupling for $L=8$, $\phi = 2\pi/L$, $\lvert\alpha\rvert^2 = 100$, $\widetilde{\Delta} = -J-\omega_m-\gamma_c/2$, $\gamma_c/\omega_m = 1\cdot 10^{-1}$, $\gamma_m/\omega_m = 1\cdot 10^{-3}$ and $\bar{n}=100$. $\delta \geq 5\%$ in the dashed region. (d) Relative error $\delta$ associated to the Fig.~\ref{fig:4} of the main text.
		}
	\end{figure}

	As shown in Fig.\,\ref{fig:5}\,(a) and (c) of the main text, and Fig.\,\ref{fig:S1}, the analytical results obtained from the effective theory match the numerical solution of the linearised dynamics in a wide regime of parameters. 
	
	\section{Additional remarks on the peculiar features of our proposed optomechanical configuration and theoretical effective description}
		
	In our letter, we have examined peculiar heat transport properties mediated by correlated fluctuations of a lattice of optical modes. We have shown that the weak coupling of initially-uncoupled modes to a common Markovian reservoir with finite correlation length, a situation relevant beyond the specific optomechanical system hereby described, induces dissipative as well as coherent processes that drastically alter the fate of the system. In particular, in a ring geometry, we identify permanent gauge currents whirling around the ring of optomechanical resonators. This result is original as the generation of these currents happens in the absence of any direct coupling between resonators and in the presence of thermal relaxation with local baths at non-cryogenic temperatures. In this scenario, the action of the extended reservoir is twofold: it builds coherence between distant thermal modes and allows for parity-breaking scattering events between distant modes. In this section, we discuss more details concerning the differences of this configuration with respect to previous interesting works in the literature.
		
	The coupling of $N$ independent mechanical modes to $N-1$ independent optical ones was considered in Refs.~\cite{Xuereb2014,Xuereb2015} and shown to allow one to generate reconfigurable interactions between distant resonators with great flexibility. In this setup, each ``reservoir'' optical mode couples to all mechanical ones with the same phase. The engineered interaction between the various mechanical modes is thus symmetric,  $\hat{H}_\mathrm{eff}=\sum_{ij}S_{ij}\hat{b}_i^\dagger\hat{b}_j^{\mathstrut}$, with $S_{ij} = S_{ji} \in \mathbb{R}$, and thus generates no gauge heat currents, which are the central point of our work.
		
	Ref.~\cite{Schmidt15} proposes two methods, namely (i) to time-modulate the coherent mechanical populations of local mechanical modes or (ii) to implement a wavelength conversion scheme in order to generate a synthetic gauge field for a lattice's photons instead of phonons. Apart from the difference in the nature of the bosonic carriers, approach (i) is completely different from our configuration that does not require modulation of the populations. Both approaches (i) and (ii) in 	Ref.~\cite{Schmidt15} would require the presence or the engineering of two-site direct couplings between the optical modes of the lattice in order to generate a photon current. In contrast, in this work no direct mechanical coupling is involved in phonon transport.
		
	In Ref.~\cite{Fazio2018}, the authors give a detailed description of persistent currents across spin chains. In that reference, in contrast to our model, the effect is achieved through a proper reservoir engineering of two-site non-reciprocal couplings \cite{Metelmann2015} between adjacent lattice sites.
		
	In Ref.~\cite{Hafezi2018}, the authors examine singular transport properties across an open chain of optomechanical resonators with a gradient of optical mean-field phases similar to the one of our letter but with nearest-neighbor coupling between both optical and mechanical modes. In that work there is a direct phonon-phonon coupling, which is responsible for the peculiar tilting of the band structure of the chain. In their setup nonreciprocal transport properties show a crucial dependence on the asymmetric gaps of the system, around which excitations have a hybrid phonon-photon nature, as opposed to the system described in our letter where the gap is not resolved ($\lvert g\alpha\rvert < \gamma_c$) and the transported excitations are of pure phononic nature.
		
	Furthermore, the theoretical method we have applied for the proposed configuration differs significantly from Refs.~\cite{Schmidt15,Fazio2018,Hafezi2018} by introducing the concept of extended reservoir. Indeed, we have first provided a general description of the effective dynamics of a set of system modes in local contact with an extended reservoir (a situation that is not restricted to optomechanics) in terms of nonlocal coherent interaction terms as well as nonlocal dissipative processes. We have shown the generality of the approach in Section~\ref{app:D} of the present SM by deriving an effective many-mode squeezing Liouvillian for initially uncoupled mechanical modes from the elimination of a two-tone driven lattice of optical cavities. This description allows us to derive the analytical expression of the current circulating around the ring as a function of the optical phase gradient and the system parameters.
	
	\section{Continuity equation in a general setting\label{app:C}}
	
	In the effective model discussed in the main text, all sites are potentially mutually coupled via the reservoir so the current operators are to be carefully defined.
	
	To do so, let us consider a lattice system as defined by some graph $G=(\Sigma,E)$ with vertices $\Sigma$, edges $E$ and one of its subsystems $G[\Omega]$ defined as the subgraph supported on some subset of vertices $\Omega \subseteq \Sigma$. Furthermore, let us denote $\hat{O}_\Omega = \sum_{i\in\Omega}\hat{O}_i$ some extensive observable $\hat{O}$. Then, $\hat{O}_\Omega$ must satisfy a continuity equation of the form
	\begin{equation}
	\partial_t\hat{O}_\Omega = -\sum_{\mathclap{(i,j)\in\partial\Omega}} \hat{\jmath}^O_{i\rightarrow j} + \hat{\sigma}_\Omega^O = \mathcal{L}^\dagger\hat{O}_\Omega,
	\end{equation}
	where the boundary $\partial\Omega$ of the subsystem $G[\Omega]$ was defined as the set of edges $\partial\Omega = \lbrace(i,j)\in E : i\in\Omega, j\notin\Omega  \rbrace$ directed from the subsystem to the rest of the system. $\hat{\sigma}_\Omega^O$ is a source term and $\mathcal{L}^\dagger$ is the adjoint Liouvillian \cite{breuer-petruccione2007} driving the operator dynamics.
	
	Now let us split the adjoint Liouillian as $\mathcal{L}^\dagger = \mathcal{L}_\mathrm{C}^{O\dagger} + \mathcal{L}_\mathrm{NC}^{O\dagger}$ into a part conserving the total amount of $\hat{O}$ ($\mathcal{L}_\mathrm{C}^{O\dagger}\hat{O}_\Sigma$) and the rest $\mathcal{L}_\mathrm{NC}^{O\dagger} = \mathcal{L}^{\dagger} - \mathcal{L}_\mathrm{C}^{O\dagger}$. By definition, $\partial\Sigma = \lbrace\varnothing\rbrace$ so the current contribution vanishes for $\Omega \rightarrow \Sigma$ ($\partial_t\hat{O}_\Sigma = 0 + \hat{\sigma}_\Sigma^O$). Therefore, $\forall \Omega \subseteq \Sigma$:
	\begin{equation}
	\mathcal{L}_\mathrm{C}^\dagger\hat{O}_\Omega = -\sum_{\mathclap{(i,j)\in\partial\Omega}}\hat{\jmath}^O_{i\rightarrow j}\;,
	\qquad
	\mathcal{L}_\mathrm{NC}^\dagger\hat{O}_\Omega = \hat{\sigma}_\Omega^O.
	\end{equation}

	For the model discussed hereby, once linear terms are absorbed into a static coherent displacement, the Louvillian conserving the total internal energy $\hat{U}_\Sigma = \sum_{i\in\Sigma}\omega_m\hat{d}_i^\dagger\hat{d}_i^{\mathstrut}$ is the effective Hamiltonian $\mathcal{L}_\mathrm{C}^{U\dagger}\placeholderdot = i[\hat{H}_m^\mathrm{eff},\placeholderdot]$ while the effective dissipator acts as a source/sink. From
	\begin{equation}
	\mathcal{L}_\mathrm{C}^{U\dagger} \hat{U}_\ell = -\sum_{\mathclap{1\leq p<L}}(\hat{\jmath}^U_{\ell\rightarrow \ell+p} + \hat{\jmath}^U_{\ell\rightarrow \ell-p}),
	\end{equation}
	one obtains the definition of heat current used in the main text $\hat{\jmath}^U_{\ell\rightarrow \ell+p} = \omega_m\hat{\jmath}_{\ell\rightarrow\ell+p} = -{\textstyle\sum_{\pm}} \frac{J_p^{\pidx{\pm}}}{2i}(\hat{d}_{\ell+p}^\dagger\hat{d}_{\ell}^{\mathstrut} e^{\mp i\phi p} - \mathrm{H.c})$.
	
	Moreover, if $\mathcal{L}^{U}_\mathrm{NC}\hat{\rho}_m(t\rightarrow+\infty) = 0$, as is the case in the main text, then $\lim_{t\rightarrow+\infty}\mathrm{Tr}[\hat{\rho}_m\mathcal{L}_\mathrm{C}^{U\dagger} \hat{U}_\ell] = 0$, i.e. either $\langle\hat{\jmath}^U_{\ell\rightarrow \ell\pm p}\rangle_{t\rightarrow\infty} = 0$ (no permanent current) or $\langle\hat{\jmath}^U_{\ell\rightarrow \ell+p}\rangle_{t\rightarrow\infty} = -\langle\hat{\jmath}^U_{\ell-p\rightarrow \ell}\rangle_{t\rightarrow\infty}$ (permanent current). To discriminate between these two cases, one can define a directional circulating current $\hat{\jmath}_{C}^U = \omega_m\sum_{\ell=1}^L\sum_{1\leq p<\lceil L/2\rceil}\hat{\jmath}_{\ell\rightarrow\ell+p}$ as done in the main text which only vanishes in the absence of permanent currents, thus serving as a witness of existence of permanent currents.
	
	\section{Effective multi-mode squeezing from a two-tone-driven extended reservoir\label{app:D}}
	
	Let us consider the general Liouvillian defined in Eqs.~\eqref{eq:1} and \eqref{eq:3} of the main text. By a two-tone driving of the cavities so as to have $\alpha_i(t) = \alpha_i^\pidx{+} e^{-i\omega_m^\pidx{i}t} + \alpha_i^\pidx{-} e^{+i\omega_m^\pidx{i}t}$ (in the frame rotating at $\omega_p$) as optical mean-field solutions, in the spirit of reference \cite{Kronwald2013}, the coupling Hamiltonian in Eq.~\eqref{eq:4} reads in the interaction picture of the mechanical modes:
	\begin{equation}
	\hat{V}_i = - \big(G_i^\pidx{+}\hat{d}_i^{\mathstrut}+G_i^\pidx{-}\hat{d}_i^\dagger\big)\hat{c}_i^{\mathstrut} + \mathrm{H.c.} - \big(G_i^\pidx{+}\hat{d}_i^{\dagger}e^{+2i\omega_m^\pidx{i} t}+G_i^\pidx{-}\hat{d}_i^{\mathstrut} e^{-2i\omega_m^\pidx{i} t}\big)\hat{c}_i^{\mathstrut} + \mathrm{H.c.},
	\end{equation}
	where subdominant terms ($\mathcal{O}(g\hat{c}^\dagger\hat{c}\hat{d})$) were neglected. As in the above reference, we define $G_i^\pidx{\pm} = g_i\alpha_i^\pidx{\pm}$ and take $\lvert G_i^\pidx{\pm}\rvert = G^\pidx{\pm}$, $\forall i$. By neglecting the rapidly varying last term and defining mechanical Bogoliubov modes $\hat{\beta}_i = \cosh r e^{i\theta_i}\hat{d}_i^{\mathstrut} + \sinh r e^{i\varphi_i}\hat{d}_i^{\dagger}$, where $r = G^\pidx{+}/G^\pidx{-}$, $\theta_i = \mathrm{Arg}(G_i^\pidx{+})$ and $\varphi_i = \mathrm{Arg}(G_i^\pidx{-})$, the effective coupling becomes $-\eta\,\hat{c}_i\hat{\beta}_i^\dagger + \mathrm{H.c.}$ with $\eta = \sqrt{G^{\pidx{-}2}-G^{\pidx{+}2}}$.
	
	By using the result of the main text, one gets the following effective Liouvillian by tracing out the optical degrees of freedom:
	\begin{align}
	\hat{H}_m^{\mathrm{eff}} &= \sum_{ij} \Omega_{ij} \hat{\beta}_i^\dagger\hat{\beta}_j^{\mathstrut},\\
	\mathcal{D}_m^{\mathrm{eff}} &= \sum_i\gamma_m^{\pidx{i}}\big((\bar{n}_i + 1)\mathcal{D}[\hat{d}_i^{\mathstrut}] + \bar{n}_i \mathcal{D}[\hat{d}_i^{\dagger}]\big)+ \sum_i\Gamma_{i} \mathcal{D}[U_{i\;\;}^{\;\;j}\hat{\beta}_j],
	\end{align}
	with
	\begin{equation}
	\mathbf{S} + \mathbf{S}^\dagger = \mathbf{U}^\dagger \mathrm{Diag}(\lbrace\Gamma_{i}\rbrace) \mathbf{U}
	\;;\quad
	\pmb{\Omega} = \frac{1}{2i}\big(\mathbf{S} - \mathbf{S}^\dagger\big)
	\end{equation}
	and
	\begin{equation}
	\mathbf{S} = -i\eta^2\mathbf{B}^{-1}
	\;;\quad
	\mathbf{B} = -\frac{J}{z}\mathbf{A} - \mathrm{Diag}(\lbrace \widetilde{\Delta}_i+i\tfrac{\gamma_c^{\pidx{i}}}{2}\rbrace).
	\end{equation}

	By defining $\alpha_{ij} = \mathrm{Arg}(\Omega_{ij})$, $\theta_{ij} = (\theta_i - \theta_j)/2$ and $\varphi_{ij} = (\varphi_i - \varphi_j)/2$, $\Theta_{ij} = (\theta_i + \theta_j)/2$ and $\Phi_{ij} = (\varphi_i + \varphi_j)/2$, one can rewrite:
	\begin{align}
	\hat{H}_m^{\mathrm{eff}} &= \sum_i\Omega_{ii}\hat{\beta}_i^\dagger\hat{\beta}_i^{\mathstrut} + \sum_{i>j} 2\lvert\Omega_{ij}\rvert \big(\sinh^2 r  \cos(\varphi_{ij}+\theta_{ij}-\alpha_{ij}) e^{i(\varphi_{ij} - \theta_{ij})} + \tfrac{1}{2}e^{i\alpha_{ij} -i(\theta_{i} - \theta_{j})}\big)\hat{d}_i^\dagger\hat{d}_j^{\mathstrut} + \mathrm{H.c.}\nonumber\\
	&+ \sum_{i>j} 2\lvert\Omega_{ij}\rvert e^{i(\Theta_{ij} - \Phi_{ij})}\cosh r\sinh r\cos(\theta_{ij}+\varphi_{ij}-\alpha_{ij})\hat{d}_i^\dagger\hat{d}_j^\dagger + \mathrm{H.c.}
	\end{align}
	
	For instance, by having $J \lessapprox \gamma_c$ so that next-to-nearest-neighbors terms can be dropped (see Fig.\,\ref{fig:S1}\,(a)) and choosing the phase of the drives so as to have $\theta_{\ell+1} - \theta_\ell = \varphi_{\ell+1} - \varphi_\ell = \alpha_{\ell+1,\ell}\; [2\pi]$ and $\theta_\ell - \varphi_\ell = \nu$, we obtain:
	\begin{equation}
	\hat{H}_m^{\mathrm{eff}} = \sum_i\Omega_{ii}\hat{\beta}_i^\dagger\hat{\beta}_i^{\mathstrut}+ \sum_{i}\big[ 2\lvert\Omega_{i+1,i}\rvert \big(\sinh^2 r + \tfrac{1}{2}\big)\hat{d}_{i+1}^\dagger\hat{d}_i^{\mathstrut} + 2\lvert\Omega_{i+1,i}\rvert \cosh r\sinh r\,e^{i\nu}\hat{d}_{i+1}^\dagger\hat{d}_i^\dagger + \mathrm{H.c.}\big].
	\end{equation}
	The system is thus subject to multi-mode squeezing as long as the system remains stable. By combining this with the engineered complex tight-binding interaction, one could in principle obtain a dissipative version of the bosonic Kitaev-Majorana described in reference \cite{McDonald2018}.
	
\end{document}